\documentclass[showpacs,prl,aps]{revtex4}
\usepackage{amsfonts}
\usepackage{amsmath}
\usepackage{amsbsy}
\usepackage{amssymb}

\font\grb=eurb10 
\def\bphi{\hbox{\grb\char'047}\,}
\def\bpsi{\hbox{\grb\char'040}\,}

\begin{document}

\title{Comment on the paper of F.J.Ernst, V.S.Manko and E.Ruiz\\ "On interrelations between Sibgatullin's and Alekseev's approaches to the construction of exact solutions of the Einstein-Maxwell equations"\\
(\rm J. Phys.: Conf. Ser. 229 (2010) 012050; arXiv:1006.5118v1)}

\author{G. A. Alekseev$^1$}
  \email{G.A.Alekseev@mi.ras.ru}
\affiliation{  \centerline{\hbox{$^1$Steklov Mathematical
Institute RAS, Moscow, Russia}}
}


\begin{abstract}
The necessity of this Comment was invoked by numerous mistakes, erroneous discussions and misleading citations curiously collected in the paper of F.J.Ernst, V.S.Manko and E.Ruiz and concerning the interrelations between two integral equation methods developed for
solution of Einstein - Maxwell equations more than twenty five years ago. At first, we clarify the origin of the errors in the paper of F.J.Ernst, V.S.Manko and E.Ruiz which gave rise to so  curious authors "conclusions" as that the monodromy transform integral equations "...are simple combinations of Sibgatullin's integral equations and normalizing conditions..." or  even that "...in the electrovac case Alekseev's integral equations are erroneous...". In the Comment, the way of correct derivation of Sibgatullin's reduction of the Hauser and Ernst integral equations in the context of the monodromy transform approach is briefly
outlined. In response to various speculations and priority claims collected in the section 3 of the F.J.Ernst, V.S.Manko and E.Ruiz paper, the concrete references are given here to the papers which were ignored completely by these authors and which show that the so called "extended electrovacuum N-soliton solutions" considered by E.Ruiz, V.S. Manko and J. Martin in 1995, are not new because all these solutions are the particular cases of a larger class of solutions constructed much earlier in explicit (determinant) form using the monodromy transform equations, and that the real story of construction of the solution for superposition of fields of two
Reissner - Nordstr\"om sources and of corresponding equilibrium configurations found in our papers with V.Belinski differs crucially from that, which one can read in the paper of
F.J.Ernst, V.S.Manko and E.Ruiz.
\end{abstract}


\maketitle

\subsection*{\large \bf On the integral equation methods and their interrelations}

\noindent\paragraph{\bf Hauser and Ernst integral equations effecting Kinnersley-Chitre transformations.} In a series of papers Kinnersley and Chitre \cite{Kinnersley:1977} -- \cite{Kinnersley-Chitre:1978b} constructed an infinite hierarchies of matrix potentials associated with each stationary axisymmetric solution of the  Einstein - Maxwell equations and showed that these  equations admit an infinite-dimensional algebra of internal symmetries which act on these potentials. An important result of Kinnersley and Chitre was also a discovery that this hierarchy of potentials (for vacuum fields) possess a generating function, which should satisfy a system of linear partial differential equations with a free complex parameter. A method for "effecting" these infinitesimal symmetry transformations was found by I.~Hauser and F.J.~Ernst in \cite{Hauser-Ernst:1979a} for vacuum case and then generalized to electrovacuum in \cite{Hauser-Ernst:1979b}. Many details of this approach were elaborated in \cite{Hauser-Ernst:1980a}-\cite{Hauser-Ernst:1980c}
These authors had shown that the generating function for the Kinnersley and Chitre matrix potentials can be determined by solving some homogeneous Hilbert problem which reduces to a system of linear singular integral equations of a Cauchy type. (It is worthy to note that emergence of the integral equations of such type is not unexpected in this context: a year before, Belinski and Zakharov\cite{Belinski-Zakharov:1978}, in the framework of their formulation of the inverse scattering method, reduced the solution of vacuum Einstein equations with two-dimensional symmetries and of equivalent spectral problem to solution of a Riemann - Hilbert problem and corresponding system of linear singular integral equations of the same type.) An important feature of the Hauser and Ernst integral equation method (similarly to the Belinski and Zakharov one) is that this represents a solution generating method: the (matrix) kernel of  the integral equations includes the functional parameters which characterize an arbitrary chosen seed and the element of the transformation group. It is important to note also that in the construction of I.~Hauser and F.J.~Ernst, the condition of regularity of the axis is used as additional simplifying assumption.
\medskip

\noindent\paragraph{\bf Sibgatullin's reduction of Hauser and Ernst singular integral equations.} A few years later, starting from
the Hauser and Ernst homogeneous Hilbert problem and corresponding matrix linear singular integral equation, N.~Sibgatullin in   \cite{Sibgatullin:1984a, Sibgatullin:1984b} refused of an arbitrary choice of a seed solution and confined his considerations by a special case of Minkowski seed.
Then, following the Hauser and Ernst assumption of minimal character of singularities of the matrix generating  function (valid only for the fields with regular axis of symmetry) and using some known technical details (see, e.g. \cite{Hauser-Ernst:1980c}), N.~Sibgatullin transformed in a standard way the integrals over the contour surrounding two singular points of generating function to the integrals over the straight cut joining these singularities. This transformation reduced the Hauser and Ernst integral equations to a decoupled system of three scalar equations (with the same scalar kernel) for three independent  functions which describe the discontinuities ("jumps") on the cut of the components of the generating function for the Kinnersley and Chitre potentials. To construct the Ernst potentials, it is sufficient to solve only one of these three equations, and in the papers of V.S.~Manko with coauthors, the simplest of these equations began to be referred to as the Sibgatullin's equation.
\medskip

\noindent\paragraph{\bf Monodromy transform approach.} The monodromy transform approach suggested by the present author in  \cite{Alekseev:1985} was developed almost at the same time as the Sibgatullin's reduction of the Hauser and Ernst integral equations. This is the most general approach to solution of Einstein - Maxwell equations for space-times with two commuting isometries. It is valid for construction not only of stationary axisymmetric fields, but also of various time-dependent fields with two commuting spatial symmetries such as colliding plane, cylindrical and some other types of waves or partially inhomogeneous cosmological models. Even in the stationary axisymmetric case,
for which the Hauser and Ernst integral equation method and its Sibgatullin's reduction were constructed only, in the monodromy transform approach, in contrast with these methods, the conditions like the regularity of the axis of symmetry as well as mathematically similar conditions of an absence of initial or focusing singularities for time-dependent fields are not imposed and therefore, many other types of solutions such as e.g., Levi-Civita and van Stockum solutions for stationary fields as well as Kazner solution and many partially inhomogeneous cosmological models or  solutions for colliding plane waves were included in a general scheme.

The basic system of linear singular integral equations, solving the inverse problem of the monodromy transform, allowed to construct infinite hierarchies of electrovacuum solutions with arbitrary  rational axis data (including asymptotically flat and asymptotically non-flat solutions) as well as solutions of non-soliton types.
Besides that, the basic for this approach concept of coordinate-independent monodromy data which characterize any solution, was introduced in such a way that these data can be determined not necessarily in terms of the data on the symmetry axis, but also on any space-time boundary (including the null ones) where the boundary data for the solution can be given. This had opened the way for solution of various boundary, initial and characteristic initial value problems for these types of fields, because the solution of the mentioned above system of singular integral equations with the monodromy data determined from given boundary data, allow to find (in principle, at least) the solution of this initial or boundary value problem.

The key points of the monodromy transform approach, such as definition of the space of normalized local solutions for the symmetry reduced Einstein - Maxwell equations, normalization conditions for the solution of the spectral problem, definition of the monodromy data functions for any local solution, derivation of the linear singular integral equations solving the inverse problem of the monodromy transform, the existence and uniqueness of solutions of these integral equations for any given set of monodromy data and others were explained in detail later -- see, for example, \cite{Alekseev:1986a} -- \cite{Alekseev:2007}. Various applications of this approach were considered in the papers \cite{Alekseev:1986b} -- \cite{Alekseev-Belinski:2007c}.

\noindent\paragraph{\bf Hauser -- Ernst -- Sibgatullin equations in the context of the monodromy transform.}
As it was already mentioned above, the basic system of linear singular integral equations solving the inverse problem of the monodromy transform  describes the entire space of local solutions of the symmetry reduced Einstein - Maxwell equations (the hyperbolic reductions as well as the elliptic ones). For the stationary axisymmetric fields, the space of solutions of these integral equations includes, besides the fields with regular axis of symmetry, all types of singular solutions which constitute a "half" of static solutions from the Weyl class and their stationary analogues. Therefore, to consider in this context the Sibgatullin's reduction of the Hauser and Ernst integral equations, we have to restrict the general monodromy transform equations
by the class of stationary axisymmetric fields with a regular axis of symmetry. This restriction reduces in half the number of arbitrary functions in generic monodromy data and implies these data to be of a special structure (analytically matched monodromy data) which admit  explicit expression in terms of the axis data for the Ernst potentials. However, even after these obvious simplifications, the monodromy transform integral equations and the Hauser -- Ernst -- Sibgatullin integral equations can not become identical, but these remain to have some important structural differences.
In particular, these equations have different kernels and different structures of the cuts on the complex planes where the integrals of the Cauchy type are defined. The expressions for the Ernst potentials in terms of solutions of these integral equations are also different.

To clarify the question a bit more, we note that the Sibgatullin's reduction of the Hauser and Ernst integral equations arise not from the reduced monodromy transform integral equations but from the equations "dual" (in some sense) to the former ones \cite{Alekseev:2005}. However, one should be careful in transforming in these dual equations a compound cut into a straight line joining a pair of singularities like in the Hauser - Ernst - Sibgatullin equations. Namely, if we merge the initial points of two parts of the cut, the integrand in the dual equations gets a pole and therefore, instead of  integral equations on two curves we obtain an integral equation on a straight line (like in the Hauser - Ernst - Sibgatullin integral equations) and the condition of vanishing of the residue at the pole mentioned above. Just these integral equations on the straight line cut and condition of vanishing residue coincide with the Hauser - Ernst - Sibgatullin integral equations and the normalization conditions for their solutions respectively.

We are not going to present here a detail expressions for such derivation  of the Hauser - Ernst - Sibgatullin integral equation and normalization condition on its solutions from the monodromy transform integral equations reduced to this class of fields (hopefully, it will be a subject of some separate publication), but instead, we mention here only that this  derivation was analyzed in detail (among many other aspects of the monodromy transform approach) by the present author, but unfortunately, this still remained to be unpublished yet. However, an interested reader can find the key points of this derivation in the  presentation of my talk entitled "Integrable reductions of Einstein's field equations: monodromy transform and the linear integral equation methods" which was given in the Isaac Newton Institute for Mathematical Sciences in the framework of the Programme "Global Problems in Mathematical Relativity", November 21, 2005 (page 22). The abstract and the presentation are still available on the website of this Institute at the addresses\\

\vskip-1ex
\qquad$<$ http://www.newton.ac.uk/programmes/GMR/alekseev.html $>$,\\[-2ex]

\qquad $<$ http://www.newton.ac.uk/webseminars/pg+ws/2005/gmr/1121/alekseev/ $>$.
\bigskip

\noindent\paragraph{\bf N-soliton solutions and the  solutions with rational axis data}
The N-soliton solutions for vacuum Einstein equations had been  discovered by Belinski and Zakharov \cite{Belinski-Zakharov:1978, Belinski-Zakharov:1979} in the framework of their formulation of the inverse scattering method. An explicit and very compact determinant  form of these solitons was constructed in \cite{Alekseev:1981}.
The N-soliton solutions of Einstein - Maxwell equations had been found in a different way \cite{Alekseev:1980a, Alekseev:1980b} using, however, the basic ideas of the inverse scattering approach. An important feature of both of these soliton generating techniques is that these admit a construction of any number of solitons on arbitrary chosen (respectively vacuum or electrovacuum) background with two commuting isometries. However, the construction of Belinski and Zakharov leads to vacuum solitons with  complex as well as real poles on the spectral plane, whereas the electrovacuum soliton generating technique \cite{Alekseev:1980a, Alekseev:1980b} admits a construction of solitons only with complex poles, while the limiting procedure, which brings a given pole to the real axis, leads in this technique to degenerate solutions with decreased number of free parameters.

A modification of electrovacuum soliton generating technique which would allow to construct electrovacuum solitons with real poles on arbitrary background has not been found. However, for electrovacuum solitons on the Minkowski background, the construction of their counterparts for which some of the poles (or all of them) are real, reduces to a purely technical problem. These solutions can be found among the solutions with rational axis data which can be calculated directly using some of the integral equation methods.

The whole class of solutions characterized by arbitrary chosen rational axis data for the Ernst potentials (or, equivalently, by arbitrary chosen rational analytically matched monodromy data) had been found in an explicit and compact (determinant) form in \cite{Alekseev:1993-GR13} using the monodromy transform integral equations (see also the Appendix in \cite{Alekseev-Garcia:1996} for more details). This class of solutions is not so large as the class of solitons in the sense, that we have no here a freedom in the choice of any  background for solitons. For  soliton solutions in this class, the background is restricted eventually by a Minkowski space-time only. On the other hand, the class of solutions \cite{Alekseev:1993-GR13} extends the class of solitons on the Minkowski background first of all because it includes, besides the asymptotically flat solutions, the none-asymptotically flat part for which the axis data functions may include, besides poles, also the polynomial parts. The asymptotically flat part of  solutions \cite{Alekseev:1993-GR13} consists of solitons on the Minkowski background (complex poles on the spectral plane) and their counterparts which have also the real poles. Thus, we can say that at least for the Minkowski background, the problem of constructing the solitons with real poles was already solved in \cite{Alekseev:1993-GR13}.
However, this is not the end of a story, and in what follows, we argue by concrete examples that
\[
\begin{minipage}{0.9\hsize}
\textit{On the Minkowski background, the soliton solutions with complex poles are joined with their counterparts (for which some poles or all of them are real) in some analytic families. For appropriate choice of parameters, different counterparts are described by the same expressions in which, however, the parameters take their values in different ranges.}
\end{minipage}
\]
We give here two examples which illustrate this situation and which occur to be very instructive concerning how to find the corresponding set of parameters for solitons.

The first example obviously is the Kerr - Newman family of solutions which includes the underextreme, the extreme and overextreme parts corresponding respectively to the fields of a black hole, extreme black hole and a naked singularity. For us it is important that this family, expressed in terms of the Boyer - Lindquist polar coordinates and physical parameters of the source (such as its mass, angular momentum, NUT-parameter, electric and magnetic charges), possess the same form for all its parts and for all values of the parameters, though for different parts of the family these parameters take their values in different ranges. On the other hand, as it is easy to show using the mentioned above soliton generating technique, the overextreme part of the Kerr - Newman family is one-soliton solution on the Minkowski background which possess a complex pole, while its underextreme part possess one real pole on the spectral plane.

Another example concerns the two-soliton solution on the Minkowski background constructed in \cite{Alekseev:1986b} (see also  \cite{Alekseev:1988} for more details), using the soliton generating technique \cite{Alekseev:1980a, Alekseev:1980b}. This 12-parametric solution describes the field of two interacting overextreme Kerr-Newman sources and possess two complex poles on the spectral plane. (More specifically, these are the poles on the spectral plane  of the fundamental solution of associated spectral problem.) Again, in this case, we can use the general form of the solutions \cite{Alekseev:1993-GR13} to find a parameterization in which all counterparts of this two-soliton solution corresponding to a pair black hole -- naked singularity  and to a pair of black holes will be described by the same analytical expressions with the only difference in the ranges of the values of this special kind of parameters.

This coalescence of different parts of two-soliton solution in a one analytical family arise even more manifestly for a static subfamily of the two-soliton case which describe the nonlinear superposition of fields of two Reissner - Nordstr\"om sources. At first, in our papers with V.Belinski \cite{Alekseev-Belinski:2007a,  Alekseev-Belinski:2007b}, we presented the 4-parametric solution which describes all equilibrium configurations of such sources. In a later paper \cite{Alekseev-Belinski:2007c}, we described in details the  intermediate steps of our construction of this solution including a complete 5-parametric static solution for the field of two interacting Reissner - Nordstr\"om sources from which these equilibrium configurations had been derived. All these solutions had been derived explicitly in a surprisingly compact form.

Actually, the 4-parametric solution for equilibrium configurations of two Reissner - Nordstr\"om sources \cite{Alekseev-Belinski:2007b} was very instructive because this allowed us to observe a remarkable fact that this solution simplify drastically if we express it in terms of the physical parameters of the sources (such as the masses calculated from Komar integrals, the charges calculated from the Gauss theorem and the distance separating the sources). The same simplification was found immediately for the 5-parametric family of solutions \cite{Alekseev-Belinski:2007c} which represents a general static subfamily of the stationary axisymmetric two-soliton solution on the Minkowski background \cite{Alekseev:1986b}. And besides that, it was found that in terms of these physical and geometrical parameters, this 5-parametric soliton solution represents an analytical family which join all counterparts corresponding to different types of the sources -- the black holes and/or naked singularities.

Thus, these examples give us the most reason to be sure that just the choice of physical and geometrical parameters of the sources as the independent parameters of various stationary axisymmetric electrovacuum soliton solutions on the Minkowski background leads to such form of these solutions which is the same for complex and real poles. This shows also that various attempts to construct some "extended soliton solutions" actually lead to the same families of  soliton solutions, provided the former are expressed in terms of an appropriate set of physically and geometrically defined free parameters.

\bigskip\noindent
This concludes our outlines of the integral equation methods and their interrelations as well as of interrelations between the classes of soliton solutions and the solutions with arbitrary axis data and now we are ready to give less formal comments of some strange aspects of the paper of F.J.Ernst, V.S.Manko and E.Ruiz and discuss  numerous erroneous points which can be found on every page of this paper.

\section*{\large  What is strange in the paper of F.J.Ernst, V.S.Manko and E.Ruiz}

\noindent{\bf N.~Sibgatullin against $<$F.J.Ernst, V.S.Manko and E.Ruiz$>$ ?}
First of all, it seems appropriate to mention that since the middle of 80th the Sibgatullin's reduction of the Hauser and Ernst integral equation method as well as the monodromy transform approach were presented and discussed by the authors on different local seminars and workshops as well as at large conferences, and during the period until 2004, when, very unfortunately, Nail Sibgatullin passed away, he had never mentioned that he had found any defects in the construction of the monodromy transform approach.

\vskip2ex
\noindent{\bf $<$F.J.Ernst, V.S.Manko and E.Ruiz$>$ against themselves?}
It seems curious that in Abstract of the paper of F.J.Ernst, V.S.Manko and E.Ruiz, one finds that
\[\left[\quad
\begin{minipage}{0.91\hsize}
\textit{The integral equations involved in Alekseev's "monodromy transform" technique are shown to be simple combinations of Sibgatullin's integral equations and normalizing conditions.}
\end{minipage}\quad
\right]
\]
On the other hand, in the next two phrases of the same Abstract one reads:
\[\left[\quad
\begin{minipage}{0.91\hsize}
\textit{An additional complex conjugation introduced by Alekseev in the integrands makes his scheme mathematically inconsistent; besides, in the electrovac case all Alekseev's principal value integrals contain an intrinsic error which has never been identified before.}
\end{minipage}\quad
\right]
\]
It is not clear,  how these equations can "contain an intrinsic error", if these are "simple combinations of Sibgatullin's integral equations  and normalizing conditions"?

\vskip3ex
\noindent{\bf $<$I.Hauser and F.J.Ernst$>$ against $<$F.J.Ernst, V.S.Manko and E.Ruiz$>$ ?}

It appears highly doubtful that Prof. Fred Ernst was actually acquainted with the material of the paper of F.J.Ernst, V.S.Manko and E.Ruiz. In 1997 our visits to the Loughborough university (UK) overlapped and we had long and friendly discussions concerning, in particular, different aspects of the Hauser and Ernst integral equation method, soliton techniques and the monodromy transform approach (with which Fred was already acquainted in details). A few years later, I.~Hauser and F.J.~Ernst published two very long papers \cite{Hauser-Ernst:1999, Hauser-Ernst:2001} where they formulated their new (generalized) homogeneous Hilbert Problem (HHP) and proved a generalized Geroch conjecture for the classes of fields described by elliptic as well as hyperbolic Ernst equations. In these papers, the authors analysed carefully the singular integral equations solving the inverse problem of the monodromy transform which were derived in  \cite{Alekseev:1985} and generalized these integral equations to the class of fields with none-smooth behaviour of solutions. Then they   used these generalized integral equations in their proofs.
To illustrate the absence of any criticism in these papers, we mentioned some titles and subtitles of the sections 2 and 3 of the I.~Hauser and F.J.~Ernst paper \cite{Hauser-Ernst:2001}:\\
\vskip-2ex
\leftskip4ex\noindent
\llap{2.} \textsc{An Alekseev-type singular integral equation that is equivalent to the HHP\\ corresponding to $(\mathbf{v}, \mathcal{F}^{M})$ when $\mathbf{v}\in \mathbf{K}^{1+}$}\\

\vskip-2ex
\leftskip5ex\noindent
2B. \textsc{Derivation of an Alekseev-type Singular Integral Equation}\\
2D. \textsc{Equivalence of the HHP to an Alekseev-type Equation When $\mathbf{v}\in \mathbf{K}^{1+}$.}\\[-2ex]

\leftskip4ex
\noindent
\llap{3.} \textsc{A Fredholm integral equation of the second kind
that is equivalent\\ to the Alekseev-type singular
integral equation when $\mathbf{v}\in \mathbf{K}^{2+}$.}\\

\vskip-2ex
\leftskip5ex\noindent
3A. \textsc{Derivation of Fredholm Equation from Alekseev-type Equation}\\

\vskip-1ex
\leftskip0ex\noindent
This contradicts obviously to the F.J.Ernst, V.S.Manko and E.Ruiz paper where in the Abstract and in its body one reads: "An additional complex conjugation introduced by Alekseev in the integrands makes his scheme mathematically inconsistent."
The only reasonable explanation of this is that Fred Ernst had not read the paper of F.J.Ernst, V.S.Manko and E.Ruiz, while V.S.Manko and E.Ruiz even had not heard about the papers of I.Hauser and F.J.Ernst mentioned just above.

\subsection*{\large \bf The origin of the errors in F.J.Ernst, V.S.Manko and E.Ruiz paper}

In the Introduction of F.J.Ernst, V.S.Manko and E.Ruiz paper, one reads:
\[\left[\quad
\begin{minipage}{0.9\hsize}
\textit{In our short communication we will show that the above equations (3)-(5) are simple combinations of Sibgatullin's equations (1) and (2).}
\end{minipage}\quad
\right]
\]
Just this attempt to identify the reduced monodromy transform equations (3)-(5) and Hauser - Ernst - Sibgatullin equations (1)-(2) gave rise in the section 2 "Derivation of Alekseev's equations" to many contradictions which urged the authors to make then such erroneous conclusions as
\[\left[\quad
\begin{minipage}{0.9\hsize}
\textit{...in the electrovac case Alekseev's integral equations (3) and (4) are erroneous...\\
...his scheme turns out mathematically inconsistent...\\
...the kernel $\mathcal{K}(\tau,\zeta)$ of all three Alekseev's equations (3) contains an intrinsic error...}
\end{minipage}\quad
\right]
\]
It is not necessary to draw the reader's attention once more to an obvious contradiction of these conclusions to the authors promise (given in the Introduction) to show that the equations (3)-(5) are simple combinations of Sibgatullin's equations (1) and (2) and to the title of this subsection 2, where one could expect to find a "derivation of Alekseev's equations". In what follows, we explain the origin
of all these erroneous statements of F.J.Ernst, V.S.Manko and E.Ruiz paper which arose appearingly from too superficial mathematical analysis of the subject.

In the first section of the present Comment we have outlined, in particular, some key points for correct derivation of the Hauser - Ernst - Sibgatullin integral equations and normalization conditions from the reduced monodromy transform equations. Here we give more details, pointing out by the way the errors in the section 2 of the paper of F.J.Ernst, V.S.Manko and E.Ruiz.

To begin with, we emphasize that the reduced monodromy transform equations (3) - (5) differ essentially from the Hauser - Ernst - Sibgatullin equations (1) - (2). In particular, the monodromy transform equations have a compound cut consisting of two disconnected parts against one straight line cut in the Hauser - Ernst - Sibgatullin equations (this fact was ignored mistakenly in the first paragraph of the section 2) and these equations do not need any supplementary normalization conditions for their solutions. Thus, these equations should not be identified with or represented as a combination of the Hauser - Ernst - Sibgatullin equations, though they describe the same space of stationary axisymmetric electrovacuum solutions with a regular axis of symmetry.

In the context of the monodromy transform approach, the Hauser - Ernst - Sibgatullin equations and normalization conditions can be obtained not from the equations (3) - (5), but from the equations which are dual (in some sense) to these equations. This means that  the column of unknown functions $\mu_a(\rho,z,\sigma)$ in the Hauser - Ernst - Sibgatullin equations should not be identified with the row of unknown functions $\bphi^a(\xi,\eta,w)$ in the monodromy transform equations:
\[\mu_a(\rho,z,\sigma)\neq\bphi^a(\xi,\eta,w=z+i\rho\,\sigma)\]
where $a=1,2,3$; $(\rho,z)$ are canonical Weyl coordinates, $\xi=z+i\rho$, $\eta=z-i\rho$, the parameter $\sigma$ runs along the straight line cut joining $w=\xi$ and $w=\eta$ on the complex plane, $w$ is a spectral parameter. Three independent  functions $\bphi^a(\xi,\eta,w)$ describe the discontinuities on the cut $L$ of the components of the matrix $\mathbf{\Psi}^{-1}(\xi,\eta,w)$ where $\mathbf{\Psi}(\xi,\eta,w)$ is the matrix of the fundamental solution of the spectral problem associated with the symmetry reduced Einstein - Maxwell equations. Instead, the functions $\mu_a(\rho,z,\sigma)$ have as the prototype the functions $\bpsi_a(\xi,\eta,w)$ which describe the discontinuities on the cut $L$ of the components of the matrix $\mathbf{\Psi}(\xi,\eta,w)$ itself:
\[\mu_a(\rho,z,\sigma)=\bpsi_a(\xi,\eta,w=z+i\rho\,\sigma)\]
In the monodromy transform approach, the integral equations for $\bpsi_a(\xi,\eta,w)$ can be considered as an alternative to the  basic system of integral equations for $\bphi^a(\xi,\eta,w)$, but the choice in favor of the equations for $\bphi^a(\xi,\eta,w)$ was made because of some technical reasons. The equations for $\bpsi_a(\xi,\eta,w)$ can be derived using the same way as the equations for $\bphi^a(\xi,\eta,w)$ and these  are defined on the same compound cut $L$ as the equations for $\bphi^a(\xi,\eta,w)$ (see \cite{Alekseev:2005} for details).

To obtain from these equations for $\bpsi_a(\xi,\eta,w)$ the Hauser - Ernst - Sibgatullin equations, we have to transform the Cauchy type integrals over a compound cut $L$ into the integrals over a straight line joining $w=\xi$ and $w=\eta$. In these transformation, it is important to take into account that after a merging of two of the four endpoints of the compound cut $L$ the integrand in each of the equations for $\bpsi_a(\xi,\eta,w)$ gets a pole outside the straight line cut joining $w=\xi$ and $w=\eta$ and each of the integral equations for $\bpsi_a(\xi,\eta,w)$ decouples into two independent equations: the integral equation on this straight line and the condition of vanishing of a residue at the pole. Just these two parts of the transformed equations for $\bpsi_a(\xi,\eta,w)$ occur to coincide identically with the Hauser - Ernst - Sibgatullin equations and normalization conditions for solutions respectively.
All this give us, in contrast with what one can read in the section 2 of the F.J.Ernst, V.S.Manko and E.Ruiz paper, an outline of a correct derivation of the Hauser - Ernst - Sibgatullin equations and normalization conditions in the context of the monodromy transform approach.

\subsection*{\large \bf Are the "extended solitons" of E.Ruiz, V.S.Manko and J.Martin\\[0.5ex] actually new and really extended?}

At the very beginning of section 3 of F.J.Ernst, V.S.Manko and E.Ruiz paper, one reads:
\[\left[\quad
\begin{minipage}{0.8\hsize}
\textit{In 1995 the extended N-soliton electrovac solution was constructed ... in a concise
analytical form [7\,${}^\prime$],...}
\end{minipage}\quad
\right]
\]
Here [7\,${}^\prime$] denotes the paper of
E.Ruiz, V.S.Manko and J.Martin, \textit{Extended N-soliton solution of the Einstein - Maxwell equations},  Phys. Rev. \textbf{D 51} 4192 (1995).

It is curious that in [7\,${}^\prime$], V.S.Manko expressed his acknowledgements to the present author "for many interesting and helpful discussions on the subject", but just in these discussions, in summer 1994, V.S.Manko was informed, in particular, about my very short but all-sufficient paper \cite{Alekseev:1993-GR13}. In this paper (and later, with more details, in the Appendix of the paper \cite{Alekseev-Garcia:1996}), all stationary axisymmetric electrovacuum solutions corresponding to arbitrary analytically matched rational monodromy data (and therefore, -- to arbitrary rational axis data for the Ernst potentials) were presented in an explicit (determinant) form.
The asymptotically flat part of solutions \cite{Alekseev:1993-GR13} includes as the particular cases \emph{all} solutions, which were presented later in [7\,${}^\prime$] as the new ones and called there as "extended solitons". However, in [7\,${}^\prime$] and in \emph{all} subsequent publications, V.S.Manko and his coauthors have ignored completely that the "extended solitons" of [7\,${}^\prime$] as well as all solutions with rational axis data published since the early 90th by N.~Sibgatullin, V.S.~Manko et al are not new, and that these solutions are the particular cases of the asymptotically flat part of the solutions  \cite{Alekseev:1993-GR13}.

Moreover, sometimes, V.S.Manko and E.Ruiz omit in the expression "extended solitons" the word "extended".
The result of this one can read, e.g., in section 3 of the paper of F.J.Ernst, V.S.Manko and E.Ruiz, in the story made up by these authors:
\[\left[\quad
\begin{minipage}{0.9\hsize}
\textit{Since the N-soliton electrovac solution ... is known, the
substitution of (14) into the formulae of the paper [7\,${}^\prime$] immediately supplies Alekseev with the explicit form of the multisoliton solution ...}
\end{minipage}\quad
\right]
\]
--- There is no need to say that I have never substituted anything into the formulae of the paper [7\,${}^\prime$]. As it was already explained in the first section of the present Comment, a convenient explicit form of a class of solutions which includes all solutions of [7\,${}^\prime$] can be found in the  earlier paper \cite{Alekseev:1993-GR13}. In concern with the beginning of the phrase above, "Since the N-soliton electrovac solution ... is known..." which creates obviously misleading impression that all electrovacuum solitons were found in [7\,${}^\prime$], we address
the reader to the first section of the present Comment where much earlier story of a discovery of vacuum and electrovacuum solitons is described in detail.

Besides that, a discussion of electrovacuum solitons on Minkowski background and the solutions \cite{Alekseev:1993-GR13} for arbitrary rational axis data given in the first section of the present Comment, shows that

(a) in view of the references \cite{Alekseev:1980a, Alekseev:1980b, Alekseev:1993-GR13}, the "extended N-soliton solitons" of E.Ruiz, V.S.Manko and J.Martin are not new, and

(b) these are not actually "extended" solitons, because for an appropriate choice of the soliton parameters, the expressions for the "extended solitons" are exactly the same as for the already known solitons on the Minkowski background, considered in a wider ranges of their parameters.

In another words, the above discussion allows us to conclude that for a special choice of parameters mentioned above the "extended solitons" simply coincide with the already known soliton solutions, provided the former are considered in a wider range of this special set of their parameters.

\subsection*{\large \bf What else is wrong in F.J.Ernst, V.S.Manko and E.Ruiz paper}
Reading of the last section 3 also brings to the readers a lot of  other misleading information which obviously needs some
comments and corrections, though this section is not closely related to the main subject mentioned in the title.

The points concerning N-soliton solutions have been discussed already in the previous section of this Comment.
Going further along the section 3 of the paper of F.J.Ernst, V.S.Manko and E.Ruiz,
we note that the transformation which was called in the title of this section as \textit{algebraic trick},
means  nothing more but a simple gauge transformation. In the monodromy transform approach, all solutions are considered in a normalized form in which all metric coefficients, field variables and Ernst potentials take their standard (Minkowski-like) values at some finite "initial" or "reference" space-time point. (Any solution can be transformed to such normalized form at any chosen regular space-time point using an appropriate gauge transformation.)
In this case, the Ernst potentials of some asymptotically flat solution normalized at some finite reference point by the conditions $\mathcal{E}=1$ and $\Phi=0$, take at spatial infinity the values which do not coincide with the Minkowski values, and pure gauge transformation (12) just provide the transformed Ernst potentials to have at spatial infinity the values $\mathcal{E}=1$ and $\Phi=0$,  i.e. it transforms these potentials to their usual asymptotically flat form.

Many other points in the section 3 of the paper of F.J.Ernst, V.S.Manko and E.Ruiz concern our paper with V.Belinski \cite{Alekseev-Belinski:2007c} which  has the number [8\,${}^\prime$] in the bibliography of   F.J.Ernst, V.S.Manko and E.Ruiz paper (we use the prime to denote the numbers from the bibliography of this paper):
\[\left[\quad
\begin{minipage}{0.9\hsize}
\textit{A careful analysis of the paper [8\,${}^\prime$] reveals,
however, that the solution generating procedure employed in [8\,${}^\prime$] reduces exclusively to rewriting
in different parameters the already known results using the following simple algebraic trick devised by Alekseev for avoiding the use of his problematic integral equations.}
\end{minipage}\quad
\right]
\]
--- "the already known results" were discussed in the previous section; for the explanation of the origin of mistakes which allowed F.J.Ernst, V.S.Manko and E.Ruiz to consider erroneously the monodromy transform equations as "problematic integral equations" we also address the readers to the previous sections of this Comment.
\[\left[\quad
\begin{minipage}{0.9\hsize}
\textit{Alekseev starts with the unphysical, asymptotically non-flat axis data...}
\end{minipage}\quad
\right]
\]
--- At first, we note that this concerns our paper with V.Belinski and therefore, it should be "Alekseev and Belinski start ...". Besides that, it is necessary to clarify that there is nothing "unphysical" in the axis data (11): these are the data for asymptotically flat solution and after a simple gauge transformation these data can be transformed to a standard asymptotically flat form.
\[\left[\quad
\begin{minipage}{0.9\hsize}
\textit{Apparently, the algebraic trick described above, if not known, causes an illusion of a true usage of the integral equations in obtaining Alekseev's results and novelty of the latter.}
\end{minipage}\quad
\right]
\]
--- It is not clear, which results the authors meant here.
The monodromy transform integral equations were really used for the construction, for example, of the class of solutions \cite{Alekseev:1993-GR13} which includes, as it was mentioned above, all "extended solitons" published later by E.Ruiz, V.S.Manko and J.Martin in [7\,${}^\prime$] (for more details see the previous sections of this Comment). The novelty of results of the present author discussed in this Comment is confirmed by concrete references.
\[\left[\quad
\begin{minipage}{0.9\hsize}
\textit{In section 2 of [8\,${}^\prime$] the above formulae (14) and (15) are used for reproducing the results of the
paper [9\,${}^\prime$] ...}
\end{minipage}\quad
\right]
\]
--- In our paper with V.Belinski [8\,${}^\prime$], we had not used (14) and (15) and we had not reproduced anybody's results. In this paper, we presented the details and intermediate steps of calculations which allowed us to find all equilibrium configurations of two Reissner - Nordstr\"om sources which were presented for the first time in our earlier paper \cite{Alekseev-Belinski:2007b}.\\

The other points of the section 3 of the paper of F.J.Ernst, V.S.Manko and E.Ruiz which surely need the corrections, are related to some assertions and unreasonable priority claims of V.S.Manko (which are almost the same as in his paper [12\,${}^\prime$]) concerning the solution for the fields of two interacting Reissner - Nordstr\"om sources constructed in our paper with V.Belinski \cite{Alekseev-Belinski:2007a}--\cite{Alekseev-Belinski:2007c}:
\[\left[\quad
\begin{minipage}{0.91\hsize}
\textit{
Turning now to the two-pole electrostatic solution considered in section 3 of [8\,${}^\prime$], it should be first of all pointed out that the solution defined by formulae (34)-(38) of that paper, contrary to the authors' claim, does not correspond to the axis data (20), (21) of [8\,${}^\prime$].}
\end{minipage}\quad
\right]
\]
--- The expressions (20), (21) in our paper [8\,${}^\prime$] determine not the axis data, but the monodromy data for the constructed solution. May be, just this mistake of F.J.Ernst, V.S.Manko and E.Ruiz is the source for their erroneous conclusion that (20),(21) does not correspond to (34)-(38) in [8\,${}^\prime$].
\[\left[\quad
\begin{minipage}{0.91\hsize}
\textit{The algebraic trick is
used this time for rewriting the Breton et al double-Reissner-Nordstrom solution [11\,${}^\prime$].
}
\end{minipage}\quad
\right]
\]
--- In our papers with V.Belinski, we have not rewrite anybody's solution. In these papers, we constructed for the first time the solution for superposition of  fields of two Reissner - Nordstr\"om sources in such compact and really explicit form that we were able then to use it for further physical analysis (in an obvious contrast with previous not successful attempts of other authors to solve this problem - see our survey in \cite{Alekseev-Belinski:2007b}).
\[\left[\quad
\begin{minipage}{0.91\hsize}
\textit{
We mention in conclusion that
formulae (34)-(38) of [8\,${}^\prime$] describe a 5-parameter representation of the BMA solution ...}
\end{minipage}\quad
\right]
\]
--- The 5-parametric solution for superposition of fields of two Reissner - Nordstr\"om sources, is called in the paper of F.J.Ernst, V.S.Manko and E.Ruiz and in V.S.Manko paper [12\,${}^\prime$] as "BMA solution" or, as in previously mentioned  phrase from this paper, as "Breton et al double-Reissner-Nordstr\"om solution". However this solution constructed pure formally in the corresponding paper [11\,${}^\prime$] of N.Breton, V.S.Manko and J.A.Aguilar-Sanchez (1998) also was not a new one because it obviously is a particular case of the class of solutions already constructed explicitly many years before, in \cite{Alekseev:1993-GR13}. As it was mentioned in the previous sections, the paper \cite{Alekseev:1993-GR13} is curiously ignored by all these authors, what allowed them to use (instead of, e.g., the words "double Reissner - Nordstr\"om solution") various expressions like  "extended solitons",  "BMA", "Breton et al", etc.
\[\left[\quad
\begin{minipage}{0.91\hsize}
\textit{
We mention in conclusion that
... a 5-parameter ... solution in physical parameters first obtained in [12\,${}^\prime$] and then just rewritten in [8\,${}^\prime$] in different coordinates.}
\end{minipage}\quad
\right]
\]
--- This is an obvious attempt to attribute to the paper [12\,${}^\prime$] of the things which had
not been originated there. This time, our paper with V.Belinski \cite{Alekseev-Belinski:2007b}
was ignored. The paper of V.S.Manko [12\,${}^\prime$] was submitted several month after
our paper \cite{Alekseev-Belinski:2007b}  was published in the same PRD journal! However,
just in the paper \cite{Alekseev-Belinski:2007b}, where we found all equilibrium configurations
of two Reissner - Nordstr\"om sources, we made at first such useful observations (already
discussed in the previous sections of this Comment) that the solutions for equilibrium
configurations of two Reissner - Nordstr\"om sources simplify drastically, if we choose the
physical parameters of the sources as the independent parameters of the solution. After that,
it was a pure technical task to express in terms of these parameters a complete 5-parametric
family of solutions for superposition of fields of two Reissner - Nordstr\"om sources, which
we obviously had derived before this and then used it in \cite{Alekseev-Belinski:2007b} for
our search of equilibrium configurations of these sources. This 5-parametric family of
solutions was published (among other details of our calculations of equilibrium
configurations) a bit later in our paper \cite{Alekseev-Belinski:2007c}. Therefore, the
F.J.Ernst, V.S.Manko and E.Ruiz version of the story that "a 5-parameter ... solution in
physical parameters first obtained in [12\,${}^\prime$] and then just rewritten in
[8\,${}^\prime$]..."  is simply unfaithful.

\subsection*{\large \bf Conclusions}
A comprehensive analysis of various approaches developed for solving vacuum Einstein equations and electrovacuum Einstein-Maxwell field equations, of interrelations between these approaches and their  applications for construction of different classes of solutions represent, without any doubts, an interesting and useful issue.

Unfortunately, a very small number of existing surveys do not provide the reader with an objective and satisfactory picture of
different mathematical aspects of construction of exact solutions --  the origins and basic ideas of different approaches,  interrelations between them, already known results, various directions of applications and perspectives.

In view of this, it would be very important to avoid a publication  of various misleading
"one-side" pure critical discussions, which does not make any constructive input into a given
scientific context, but which can bring  a lot of distorted information to the  readers who would
like to learn more about the story of the subject and about the present knowledge in this
scientific area.

\subsection*{\large \bf Acknowledgements}
This work is supported in parts by the Russian Foundation for Basic Research
(grants 08-01-00501, 08-01-00618, 09-01-92433-CE) and the program ``Mathematical
Methods of Nonlinear Dynamics" of the Russian Academy of Sciences.

\end{document}